%% file: oss2017.tex
\documentclass{llncs}

\input{config}

\begin{document}

    \title{Assessing Code Authorship: The Case of the Linux Kernel}

\author{Guilherme Avelino\inst{1,2} (0000-0002-8203-0638) \and Leonardo Passos\inst{3} (0000-0001-6591-993X) \and
    Andre Hora\inst{1} (0000-0003-4900-1330) \and Marco Tulio Valente\inst{1} (0000-0002-8180-7548)}
    
    \institute{Federal University of Minas Gerais (UFMG), Brazil\\
               \email{\{gaa,mtov,hora\}@dcc.ufmg.br}
    \and
    Federal University of Piaui (UFPI), Brazil
    \and
    University of Waterloo, Canada\\
    \email{lpassos@gsd.uwaterloo.ca}
    }
    
    
    \maketitle
    
    
    \input{abstract}

    
    
    \input{introduction}

    
    \input{design}

    \input{results}

    
    
    
    \input{threats}

    
    \input{related}

    
    \input{conclusion}

  
    %

    \bibliographystyle{splncs03}
    \bibliography{references}

\end{document}

%% file: config.tex
\usepackage{cite}

\usepackage[utf8, latin1]{inputenc}
\usepackage[T1]{fontenc}

\usepackage{flushend}

\usepackage{multicol}
\usepackage{multirow}
\usepackage{graphicx}
\usepackage{framed}

\usepackage[hyphens]{url}
\usepackage[anythingbreaks]{breakurl}

\usepackage{amsmath}

\usepackage{booktabs}
\usepackage{color}
\usepackage{array}

\usepackage[caption=false]{subfig}

\newcommand{\aspas}[1]{{``#1''}}
\newcommand{\mcode}[1]{$\tt #1$}

\definecolor{gray}{RGB}{90,90,90}

\expandafter\def\expandafter\UrlBreaks\expandafter{\UrlBreaks
    \do\a\do\b\do\c\do\d\do\e\do\f\do\g\do\h\do\i\do\j%
    \do\k\do\l\do\m\do\n\do\o\do\p\do\q\do\r\do\s\do\t%
    \do\u\do\v\do\w\do\x\do\y\do\z\do\A\do\B\do\C\do\D%
    \do\E\do\F\do\G\do\H\do\I\do\J\do\K\do\L\do\M\do\N%
    \do\O\do\P\do\Q\do\R\do\S\do\T\do\U\do\V\do\W\do\X%
    \do\Y\do\Z}

%% file: abstract.tex
\begin{abstract}
Code authorship is a key information in large-scale open-source systems.
Among others, it allows maintainers to assess division of work and identify 
key collaborators. 
Interestingly, open-source 
communities lack guidelines on how to manage authorship. 
This could be mitigated by setting to build an empirical body of 
knowledge on how authorship-related measures evolve in successful open-source 
communities.
Towards that direction, we perform a case study on the Linux kernel. 
Our results show that:
(a) only a small portion of developers (26\,\%) makes significant contributions 
to the code base; 
(b) the distribution of the number of files per author is highly skewed---a small group of top-authors (3\,\%) is responsible for hundreds of files, while most authors (75\,\%) are responsible for at most 11 files; 
(c) most authors (62\,\%) have a specialist profile; 
(d) authors with a 
high number of co-authorship
connections tend to collaborate with others with less connections. 
\end{abstract}

%% file: introduction.tex
\section{Introduction}

Collaborative work  and modularization are key players in 
software development, specially in the context of open-source systems~\cite{Herbsleb2007,Mistrik2010,Parnas1972}.
In a collaborative setup imposed by open-source development, 
code authorship allows developers to identify which project members to 
contact upon maintaining existing parts of the code base. 
Additionally, authorship information allows maintainers to assess overall 
division of work among project members (e.g., to seek better working balance) and 
identify profiles within the team  (e.g., 
specialists versus generalists).

Our notion of authorship is broader than the English definition of the word. 
In the context of code, authorship relates to those who make significant 
changes to a target file. This may include the original file creator, as well 
as those who subsequently change it. 
Hence, different from authorship in books and scientific 
papers, code authorship is inherently 
dynamic as a software evolves. 

\noindent\textbf{Problem Statement.} Currently, open-source communities lack 
guidance on how to organize and manage code authorship among its contributors. 
\\[-.2cm]

\noindent\textbf{Research Goal.} We argue that the stated problem could 
be mitigated by setting to build an empirical body of 
knowledge on how authorship-related measures evolve in successful open-source 
communities.
In that direction, we investigate authorship in a large and 
long-lived system---the Linux kernel. Our goal is to identify 
authorship parameters from the Linux kernel evolution history, as well as 
interpret why they appear as such. 
We also check whether those 
parameters apply to the subsystem level, allowing us to assess their generality 
across different parts of the kernel.
Our analysis accounts for 56 stable releases (v2.6.12--v4.7), spanning a period 
of over 11 years of development (June, 2005--July, 2016). 
\\[-.2cm]

\noindent\textbf{Research Questions.} When investigating the Linux kernel 
authorship history, we follow three research questions:
\\[-.2cm]


\noindent \emph{RQ1: What is the distribution of the number of files per
    author?} \\
\emph{\underline{Motivation:}} Answering such a question provides us with a 
measure of the work overload and the concentration of knowledge within team 
members, as well as how that evolves over time.
%
\\[-.2cm]

\noindent \emph{RQ2: How specialized is the work of Linux authors?} \\
\emph{\underline{Motivation:}} Following the Linux kernel architectural 
decomposition, we seek to understand the proportion of developers who have a 
narrower understanding of the system (specialists), versus those with a broader 
knowledge (generalists). Specialist developers author files in a single 
subsystem; generalists, in turn, author files in different subsystems. 
Answering our research question seeks to assess how effective the
Linux kernel architectural decomposition is in fostering specialized work, a 
benefit usually expected from a good 
modularization design~\cite{sullivan-dsm,designrules}.
%
\\[-.2cm]

\noindent \emph{RQ3: What are the properties of the Linux co-authorship
    network?} \\
\emph{\underline{Motivation:}} The authorship metric we use enables identifying
multiple authors per file, evidencing a co-authorship collaboration among 
developers~\cite{Meneely2011}. Such collaborations form a network---vertices 
denote authors and edges connect authors sharing common authored files.
This question seeks to identify collaboration properties in the 
Linux kernel co-authorship network. 
\\[-.2cm]

\noindent\textbf{Contributions.} From the investigation we conduct, 
we claim two major contributions:
(a) an in-depth investigation of
authorship in a large, successful,  and long-lived open-source community, 
backed up by 
several
authorship measures when answering each of our research questions. 
The findings we report also serve researchers, 
allowing them to contrast the authorship in the Linux kernel  with 
those of other communities; 
(b) the definition of several authorship-centric concepts, such as authors and 
specialists/generalists, that others may use as a common ground to study the social organization 
of software systems. 

In Section~\ref{sec:study_design}, we provide a description of our study 
design. Section \ref{sec:authorship_results} details our 
results. Sections~\ref{sec:threats} 
and~\ref{sec:related_work} discuss threats
to validity and related work, respectively. Section~\ref{sec:conclusion} 
concludes the paper, also outlining future work.

%% file: design.tex
\section{Study Design}\label{sec:study_design}

\subsection{Author Identification}
\label{sec:authorship_measure}

At the core of our study lies the ability to identify and quantify authorship 
at the source code level. To identify file authors, as required by our three 
research questions, we employ a normalized version of the {\em degree-of-authorship}
(DOA) metric~\cite{Fritz2010, Fritz2014}. The metric is originally defined 
in absolute terms:\\[-.2cm]
\[
\label{eq:DOA}
\begin{aligned}
\mathit{DOA_A}(d,f) ={} & 3.293 + 1.098 * \mathit{FA} + 
   0.164 * \mathit{DL} - 0.321 * \ln(1+ \mathit{AC})
\end{aligned}
\]
From the provided formula, the absolute degree of authorship of a developer $d$ 
in a file $f$
depends on three factors: first authorship (FA), number of deliveries (DL), 
and number of acceptances (AC). If $d$ is the creator of $f$, $\mathit{FA}$ is
1; otherwise it is 0; $\mathit{DL}$ is the number of changes in $f$ made by $d$;
and $\mathit{AC}$ is the number of changes in $f$ made by other developers. 
$\mathit{DOA_A}$ assumes that FA is by far the strongest predictor of file
authorship. Further changes by $d$ (DL) also contributes positively to his authorship,
but with less importance. 
Finally, changes by other developers (AC) contribute to decrease someone's 
$\mathit{DOA_A}$, but at a slower rate. The weights we choose stem 
from an experiment with professional Java developers~\cite{Fritz2014}. 
We reuse such thresholds without further modification. 

The normalized DOA  ($\mathit{DOA}_N$) is as given in~\cite{Avelino2016}:\\[-.2cm]
\[
\begin{aligned}
\mathit{DOA_N}(d,f) ={} & \mathit{DOA_A}(d,f) / \mathit{max}(\{\mathit{DOA_A}(d',f)~|~d'\in\mathit{changed}(f)\})
\end{aligned}
\]
\noindent In the above equation, $\mathit{changed}(f)$ denotes the set of 
developers who edited a file $f$ up to a snapshot of interest (e.g., release). 
This includes the developer who creates $f$, as well as all those who 
later modify the file. $\mathit{DOA_N} \in [0..1]$: $1$ is granted to the 
developer(s) with the highest absolute DOA among those changing $f$; in 
other cases, $\mathit{DOA_N}$ is less than one. 

Lastly, the set of authors of a file $f$ is given by:\\[-.2cm]
\[
\mathit{authors}(f) = \cup \{d~|~d \in \mathit{changed}(f)\land 
\mathit{DOA_N}(d,f) > 0.75 \land \mathit{DOA_A}(d,f) \ge 3.293 \}
\]

The authors identification depends on specific thresholds--- $0.75$ and $3.293$. Those 
stem from a calibration setup when applying $\mathit{DOA_N}$ to a large corpus 
of open-source systems. For full details, we refer readers 
to~\cite{Avelino2016}. 

\subsection{Linux Kernel Architectural Decomposition}
Investigating authorship at the subsystem level requires a reference 
architecture of the Linux kernel.
Structurally, the Linux kernel architectural decomposition comprises seven 
major subsystems~\cite{Corbet2005}: 
\emph{Arch} (architecture dependent code), 
\emph{Core} (scheduler, IPC, memory management, etc),
\emph{Driver} (device drivers), 
\emph{Firmware} (firmware required by device drivers), 
\emph{Fs} (file systems), 
\emph{Net} (network stack implementation), 
and \emph{Misc} (miscellaneous files, including documentation, samples, scripts,
etc). 
To map files in each subsystem, we rely on mapping rules
set by G.~Kroah-Hartman, one of the main Linux kernel developers.\footnote{\url{
https://github.com/gregkh/kernel-history/blob/master/scripts/stats.pl}}
Table~\ref{tab:subsystems} shows the number of files in each kernel subsystem as mapped by using the expert rules. 
%
\begin{table}[t]
	\centering
	\caption{Linux subsystems size and authors proportion}
	\label{tab:subsystems}
	\begin{tabular}{crrrrrc}
		\hline
		\textbf{}          & \multicolumn{2}{|c|}{\textbf{Files}}                                & \multicolumn{4}{c}{\textbf{Authors Proportion}}                                                                                                       \\
		\textbf{}          & \multicolumn{2}{|c|}{\textbf{}}     & \multicolumn{3}{c|}{\textbf{Last release (v4.7)}}                                                                         & \textbf{All releases}      \\ 
		\textbf{Subsystem} & \multicolumn{1}{|c}{\textbf{\#}} & \multicolumn{1}{c|}{\textbf{\%}} & \multicolumn{1}{c}{\textbf{Developers}} & \multicolumn{1}{c}{\textbf{Authors}} & \multicolumn{1}{c  |}{\textbf{Proportion}} & \textbf{Avg $\pm$ Std Dev} \\ \hline
		\textbf{Driver}    & 22,943                          & 42\,\%                             & 10,771                                  & 2,604                                & 24\,\%                                    & 25.00 $\pm$ 0.80\,\%         \\
		\textbf{Arch}      & 17,069                          & 32\,\%                             & 3,613                                   & 1,145                                & 32\,\%                                    & 33.10 $\pm$ 1.28\,\%         \\
		\textbf{Misc}      & 6,621                           & 12\,\%                             & 644                                     & 78                                   & 12\,\%                                    & 14.85 $\pm$ 2.69\,\%         \\
		\textbf{Core}      & 3,840                           & 7\,\%                              & 4,165                                   & 1,083                                & 26\,\%                                    & 25.77 $\pm$ 1.56\,\%         \\
		\textbf{Net}       & 1,957                           & 4\,\%                              & 2,161                                   & 269                                  & 13\,\%                                    & 13.63 $\pm$ 0.90\,\%         \\
		\textbf{Fs}        & 1,809                           & 3\,\%                              & 1,777                                   & 175                                  & 10\,\%                                    & 12.61 $\pm$ 1.95\,\%         \\
		\textbf{Firmware}  & 151                             & 0\,\%                              & -                                       & -                                    & -                                       & -                          \\ \hline
		\textbf{All}       & 54,400                          & 100\,\%                            & 13,436                                  & 3,459                                & 26\,\%                                    & 26.86 $\pm$ 0.83\,\%         \\ \hline
	\end{tabular}
\end{table}

\subsection{Data Collection}\label{sec:dataset}

We study 56 stable releases of the Linux kernel, obtained from {\sc linus/torvalds} 
GitHub repository. A stable 
release is any named tag snapshot whose identifier does not have a \emph{-rc} 
suffix. To define the $\mathit{authors}$ set of a file $f$ in a given release 
$r$, we calculate $\mathit{DOA_N}$ from the first commit 
up to $r$. 
It happens, however, that the Linux kernel history is not fully stored under 
Git, as explained by Linus Torvalds in the first commit message.\footnote{\burl{
https://github.com/torvalds/linux/commit/1da177e4c3f41524e886b7f1b8a0c1fc7321cac2}}
Therefore, we use 
\mcode{git~graft} to join the history of all releases prior to v2.6.12 (the first 
release recorded in Git) with those already controlled by Git. 
After join, we increment the Linux kernel Git history with  64,468 
additional commits.

Given the entire development history, we check out each stable release at a 
time, listing its files, and calculating their
$\mathit{DOA_N}$. In the latter case, we rely on 
\mcode{git~log} \--\--\mcode{no}\mcode\--\mcode{merges} to discard merge commits and retrieve all the changes to a given file prior to the release 
under investigation. To compute the $\mathit{DOA_N}$, we only consider the author of each commit, not its committer (Git repositories store both)~\cite{chacon2014pro}.
It is worth noting that prior to calculate $\mathit{DOA_N}$, we map possible  
aliases among developers, as well as eliminate unrelated 
source code files. 
As example, the \emph{Firmware}
subsystem was removed because most of its files are binary blobs.
To perform these steps, we adopt the procedures described in~\cite{Avelino2016}. 

Table~\ref{tab:subsystems} shows the proportion of authors in each Linux subsystem. In the last release, Linux kernel has 13,436 developers, but only 3,459 (26\,\%) are authors of at least one file. 
Throughout the kernel development, the proportion of authors is nearly constant (\emph{Std dev}$ = \pm$ 0.83\,\%). Thus, the heavy-load Linux kernel maintenance has been kept in the hands of less than a third of all developers.

Using custom-made scripts, we fully automate authorship identification, as well 
as the collection of supporting data for the claims we make. 
Our infrastructure is publicly available on 
GitHub.\footnote{\url{https://github.com/gavelino/data\_oss17}}
We encourage others to use it as means to independently replicate/validate our 
results.

%% file: results.tex
\section{Results}\label{sec:authorship_results}
\subsection*{RQ.1) Distribution of the Number of Files per
    Author}\label{sec:numbers-of-files}
\noindent\textit{What is the distribution of the number of files per 
author?}\\[-.2cm]

\begin{figure}[!t]
  \centering  
  \includegraphics[width=\linewidth]
  {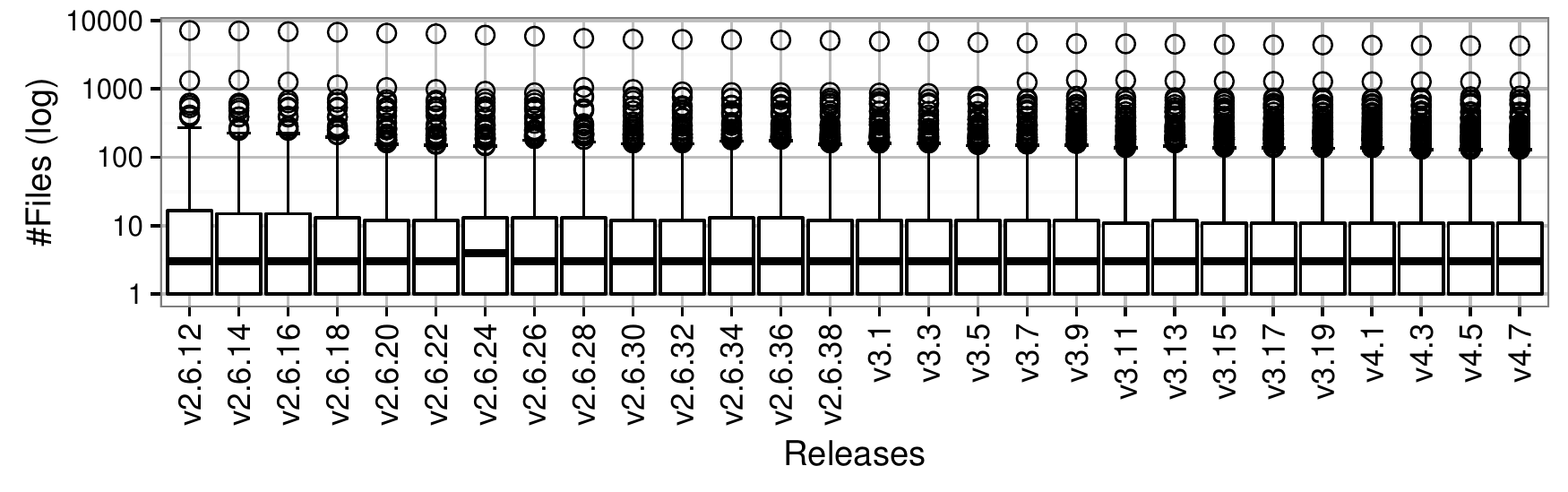}
  \caption{Distribution of the number of files per author in each release}
  \label{fig:workload-boxplot}
\end{figure}

The number of files per author is highly skewed. 
Figure~\ref{fig:workload-boxplot}  presents the boxplots of files per 
author across the Linux kernel releases (adjusted for skewness---see~\cite{Hubert2008}). To simplify the visualization of the results, we present the boxplots at each two releases.  
With exception of one release (v2.6.24),
50\,\% of the authors responds to at most three files (median); for 75\,\% of the authors, 
the number of files ranges from 11 to 16. Outliers follow from the skewed 
distribution. Still, the number of authors with more than 100 files is always 
lower than 7\,\% of the authors, ranging from 7\,\% in the first release to 3\,\% in 
the last one. Similar behavior is observed at the subsystem level. 
In the last release (v4.7), for instance, the number of files 
per author up to the 75\,\% percentile in \emph{Fs}, \emph{Arch}, and \emph{Driver} closely resemble one-another and the 
global distribution as a whole---all share the same median (three). \emph{Core} and 
\emph{Misc}, however, have less variability than the other subsystems, 
as well as lower median values (two and one, respectively).

\begin{figure}[!t]
    
    \centering
    \includegraphics[width=\linewidth]{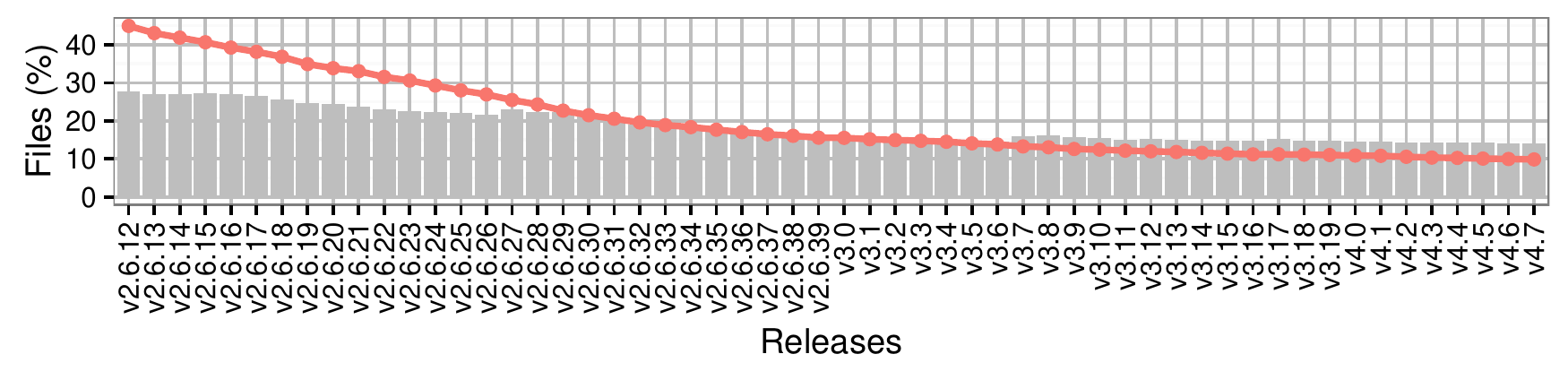}
    \caption{Percentage of files authored by the top-10 authors over time. The 
line represents Linus Torvalds (top-1) and the bars represent the accumulated 
number of files of the next top-9 authors}
    \label{fig:authorship-onefigure}
\end{figure}

It is interesting to note that file authorship  follows a pyramid-like 
shape of increasing authority; at the top, Linus Torvalds acts as a "dictator", 
centralizing authorship of most of the files (after all, he did create the 
kernel!). Bellow him lies his hand-picked "lieutenants", often chosen on the 
basis of merit. Such organization directly reflects the Linux kernel 
contribution dynamics, which is itself a pyramid~\cite{Bettenburg2015}. 
However, as the kernel evolves, we see that Torvalds is becoming more "benevolent". 
As Figure~\ref{fig:authorship-onefigure} shows,  the percentage of files 
authored by him has reduced from 45\,\% (first release) 
to 10\,\% in v4.7. Currently, he spends more time verifying and integrating 
patches than writing code~\cite{Corbet2013}.
Similar behavior is observed 
downwards the authorship pyramid. The 
percentage of files in the hand of the next top-9 Linux kernel authors (bars) is 
consistently decreasing. This suggests that authorship is increasing at lower 
levels of the pyramid, becoming more decentralized. 
This is indeed expected and 
to an extent required to allow the Linux kernel evolves at the 
pace it does.

\begin{figure}[!t]
    \centering
    \includegraphics[width=\linewidth]{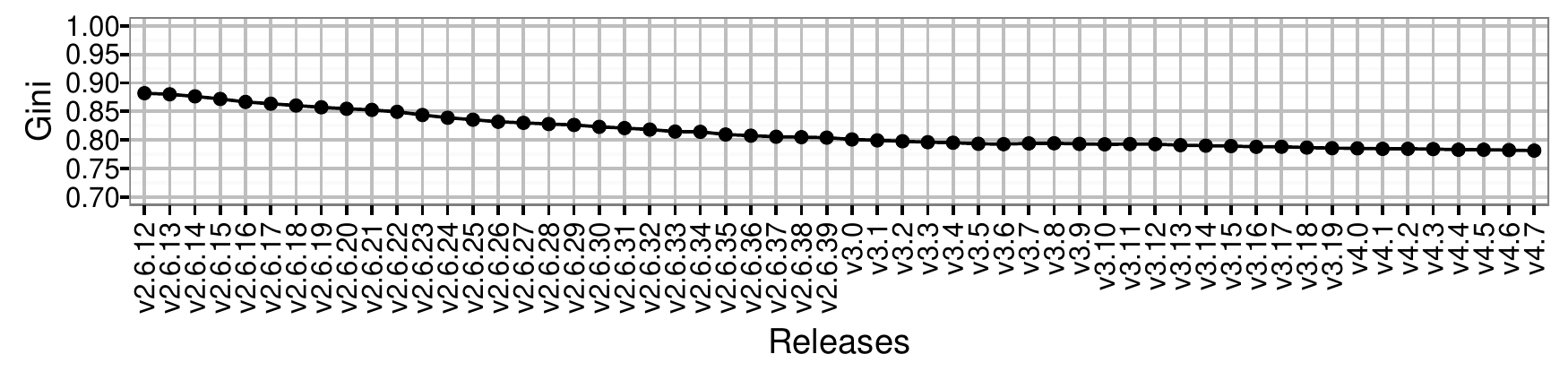}
    \caption{Gini coefficients. It ranges from 0
    	(perfect equality) to 1 (perfect
    	inequality).}
    \label{fig:gini-boxplot}
\end{figure}

We also apply the Gini coefficients~\cite{Gini} to analyze the distribution of the number of files per author (Figure~\ref{fig:gini-boxplot}). 
In all releases, the coefficient is high, confirming skewness. However, we 
notice a decreasing trend, ranging from 0.88 in the first release to 
0.78 (v4.7). Such a trend further strengthens our notion that authorship in the 
Linux kernel is becoming less centralized. 
   
\subsection*{RQ.2) Work Specialization}
\noindent\textit{How specialized is the work of Linux authors?}\\[-.2cm]

\begin{figure}[!t]
    \centering
    \includegraphics[width=\linewidth]{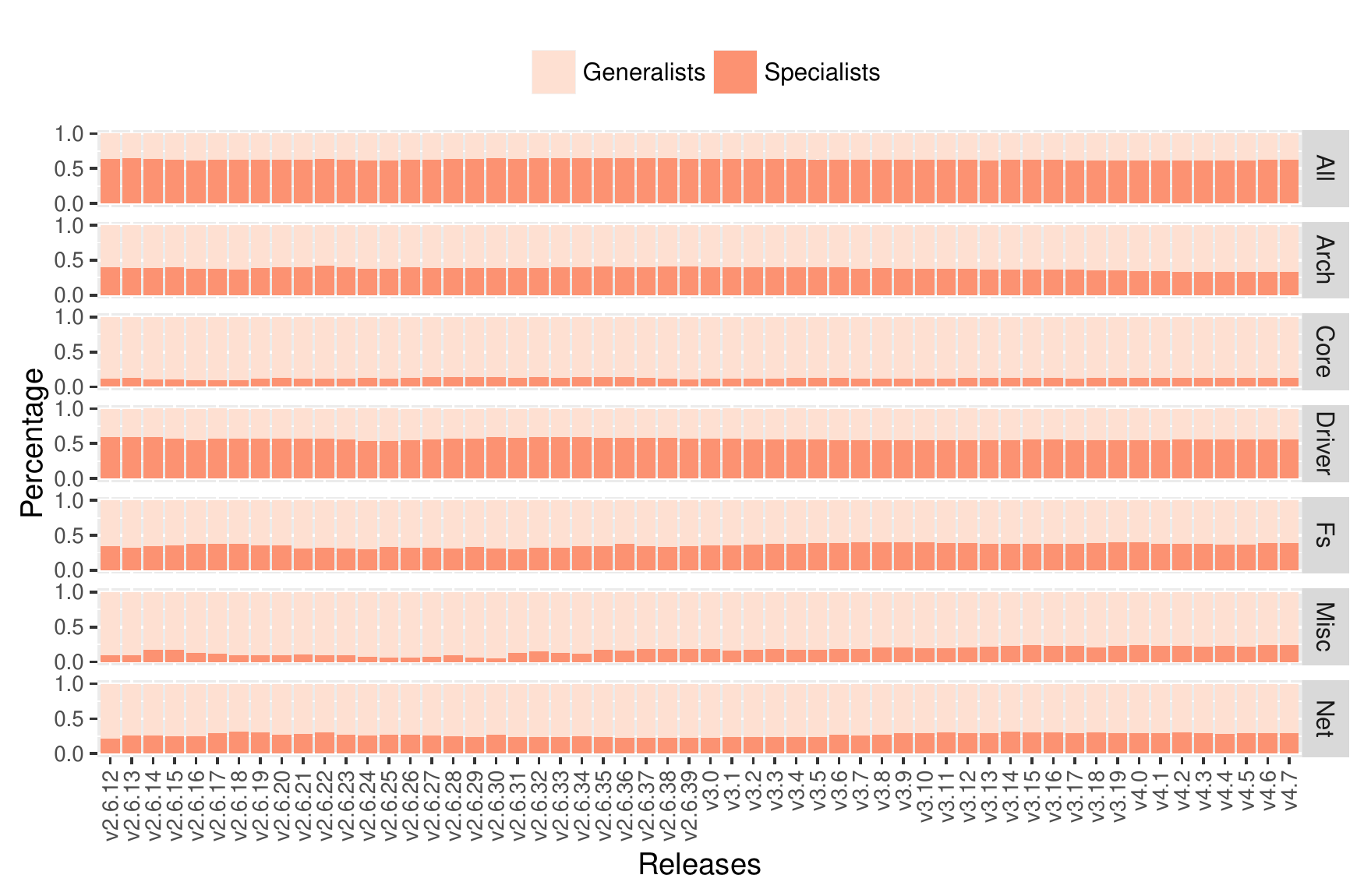}
    \caption{Percentage of specialists and generalists}
    \label{fig:spec}
\end{figure}

To assess work specialization, we introduce two author profiles.
We call authors \emph{specialists} if they author files in a single subsystem.
\textit{Generalists}, in turn, author files in at least two 
subsystems. As Figure~\ref{fig:spec} shows, the number of 
specialists dominates the amount of generalists. 
In the Linux kernel (All), any given release has at least 61\,\% of specialist authors, 
with a maximum of 64\,\%; at all times, 39\,\% of the authors are generalists.
Moreover, the proportion of generalists and specialists appears to be fairly 
stable across the entire kernel (All) and its constituent subsystems (except for
\emph{Misc}).

Looking at the work specialization in each subsystem also provides a means to 
assess how much the Linux kernel architectural decomposition fosters 
specialized work. 
The architectural decomposition plays a key role in fostering 
specialists inside the \emph{Driver} subsystem (more than 50\,\% of specialists), but less so elsewhere. 
The reason it occurs so extensively inside \emph{Driver} follows from the plug-in interface
of the latter and its relative high independence to other subsystems~\cite{Corbet2005,Passos2015}. 
In contrast, \emph{Core} and \emph{Misc} have the lowest percentage of
specialized workers. More than 75\,\% of their authors own files
in more than one subsystem. Specifically, \emph{Core} is the subsystem with the lowest percentage of specialized workers (13\,\%). This is also expected since \emph{Core} developers tend
to have expertise on Linux's central features, which allows them to also work on other subsystems.

%

\subsection*{RQ.3) Co-authorship Properties}
\noindent\textit{What are the properties of the Linux co-authorship 
network?}\\[-.2cm]

Many files in the Linux kernel result from the work of different authors. 
As such, we set to investigate such collaboration by means of the properties of 
the Linux kernel \textit{co-authorship network}. We model the latter as 
follows: vertices stand for Linux kernel authors; an edge connects two authors 
$v_i$ and $v_j$ if $\exists f$ such that 
$\{v_i, v_j\} \subseteq \mathit{authors}(f)$. In other words, an edge represents a
\textit{collaboration}. 

To answer our research question, we initially analyze 
the latest co-authorship network, as given in release v4.7 
(Table~\ref{tab:coauthorship}).\footnote{We use the R \emph{igraph} (version 1.0.1) to calculate 
all measures.}
The number of vertices (authors) determines the size of a co-authorship 
network. The mean degree network, in turn, inspects the number of co-authors 
that a given author connects to. In the system level 
(All), the mean vertex degree is 3.64, i.e., on average, a Linux author collaborates with 3.64 other authors. At the subsystem level, \emph{Driver} forms 
the largest network (2,604 authors, 75\,\%), whereas \emph{Misc} results in the 
smallest one (78 authors, 2\,\%). \emph{Arch} has the highest mean degree 
(3.14 collaborators per author); \emph{Misc} has the lowest (0.79). Linus Torvalds has connections with 215 other authors. His 
collaborations spread over all subsystems and range from 92 collaborations in 
\emph{Driver} to five in \emph{Misc}. Excluded Torvalds, the top-2 
and top-3 authors with more collaborators have 156 and 118 collaborators, 
respectively.  

\begin{table*}[tp]
	\centering
	\caption{Co-authorship network properties (release v4.7)}
	\label{tab:coauthorship}
	\begin{tabular}{lrrrrrrr}
		\toprule
		& \textbf{All}    & \multicolumn{1}{l}{\textbf{Driver}} &
		\multicolumn{1}{l}{\textbf{Arch}} & \multicolumn{1}{l}{\textbf{Core}} &
		\multicolumn{1}{l}{\textbf{Net}}    & \multicolumn{1}{l}{\textbf{Fs}} &
		\multicolumn{1}{l}{\textbf{Misc}} \\ 
		\midrule
		\textbf{Number of Vertices}           & 3,459  & 2,604                      &
		1,145                    & 1,083                    & 269    & 175              
		& 78                       \\
		\textbf{Mean Degree}               & 3.64   & 2.74                       &
		3.14                     & 1.67                     & 2.57   & 2.59             
		& 0.79                     \\
		\textbf{Clustering Coefficient}    & 0.080  & 0.074                      &
		0.128                    & 0.074                    & 0.205  & 0.175            
		& 0.188                    \\
		\textbf{Assortativity Coefficient} & -0.070 & -0.115                     &
		-0.060                   & -0.072                   & -0.003 & -0.146           
		& -0.062                   \\ 
		\hline
	\end{tabular}
\end{table*}

The third property, {\em clustering coefficient}, 
reveals the degree to which adjacent vertices of a given vertex tend to be
connected~\cite{Watts1998}. In a co-authorship network, the coefficient gives
the probability that two authors who have a co-author in common are also
co-authors themselves. 
A high coefficient indicates that the vertices tend to
form high density clusters. The clustering coefficient of the 
Linux kernel is small ($0.080$). Nonetheless, \emph{Net}, \emph{Misc}, and \emph{Fs} 
 exhibit a higher tendency to form density clusters ($0.205$, $0.188$,
and $0.175$, respectively) in comparison to other subsystems. 
The three subsystems are the smallest we analyze, a factor that influences the 
development of collaboration clusters~\cite{Barabasi2002a}. 

Last, but not least, we compute the \emph{assortativity 
coefficient}, which correlates the number of co-authors of an author (i.e., its 
vertex degree) with the number of co-authors of the authors it is connected 
to~\cite{Newman2003}. Ranging from -1 to 1, the coefficient shows whether 
authors with many co-authors tend to collaborate with other highly-connected 
authors (positive correlation). In v4.7, all subsystems have negative 
assortativity coefficients, ranging from $-0.134$ in \emph{Fs} to 
$-0.029$ in \emph{Net} subsystem.  This result diverges from the one commonly 
observed in scientific communities~\cite{Newman2004}. Essentially, this 
suggests that Linux kernel developers often divide work 
among experts who help less expert ones. These
experts (i.e., highly-connected vertices), in turn, usually do not collaborate 
among themselves (i.e., the networks have negative \emph{assortative 
coefficients}).  

We identify in the co-authorship networks a relevant amount of 
\emph{solitary authors}---authors that do not have co-authorship with any 
other developer. In total, 20\,\% (699) of Linux kernel 
developers are solitary. Although there is a high percentage of solitary authors, only 9\,\% of them have more than three files. Additionally, 66\,\% of them work in the \emph{Driver} subsystem. 
The latter is likely to follow from the high proportion of specialists within that subsystem (see RQ.2). 
%

\begin{figure}[!t]
    \centering
    \includegraphics[width=\linewidth]{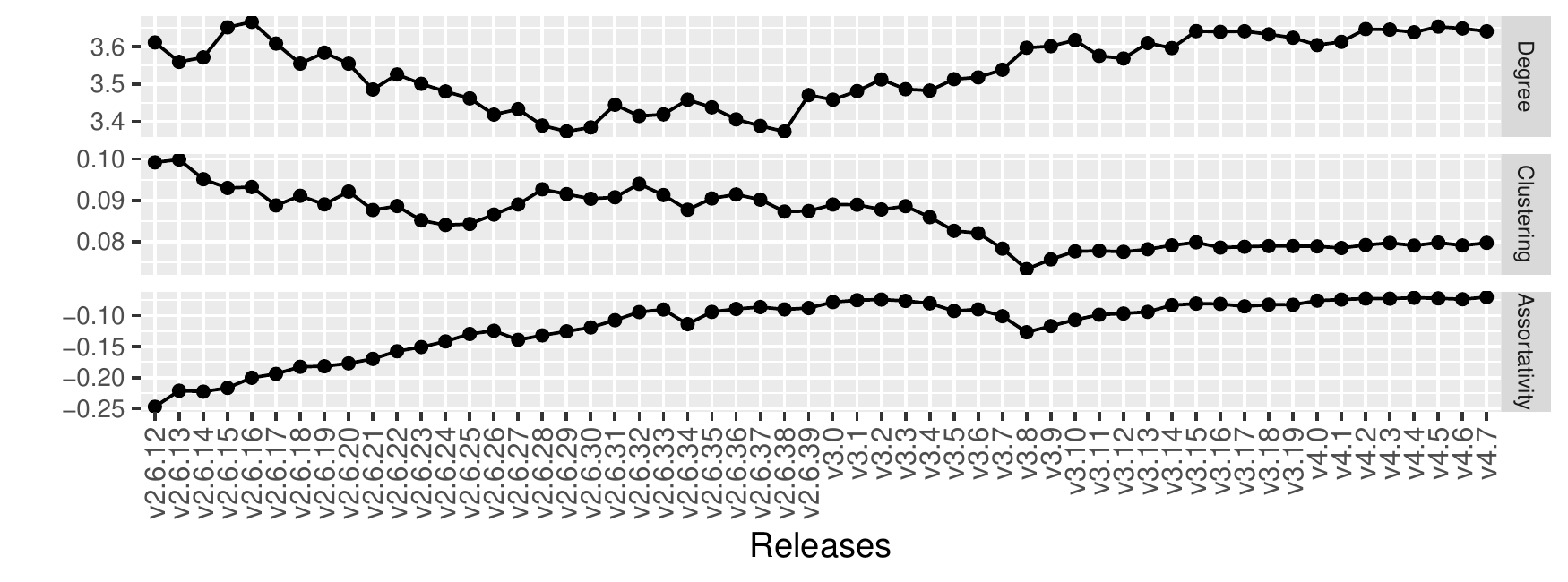}
    \caption{Co-authorship network properties over time}
    \label{fig:network-stats}
\end{figure}

\vspace{.2cm}
\noindent\textbf{Evolution of Co-authorship network properties.} 
We  set to
investigate how the co-authorship properties evolved to those in release v4.7. 
Figure~\ref{fig:network-stats} 
displays the corresponding graphics. Although we can observe a small decrease in some intermediate releases, by looking at the first and last releases, the mean degree has little variation, ranging from 3.61 to 3.64. Clustering coefficient, in turn, varies from 0.099 (first 
release) to 0.080 (v4.7). Since the mean degree does not vary 
considerably, we interpret such decrease as an
effect of the growth of the number of authors (network vertices). The latter
creates new opportunities of collaboration, but these new connections do
not increase the density of the already existing clusters. 
A similar behavior is common in other networks, as described by Albert and 
Barab\'asi~\cite{Barabasi2002a}. 
Finally, we observe a relevant variation in the evolution of assortativity 
coefficients. Measurements range
from -0.25 in the first release to -0.07 in v4.7. Such a trend aligns 
with the decrease of the percentage of files authored by Linus Torvalds and the 
other top authors (refer to RQ.1). With 
less files, these authors are missing some of their connections and becoming 
more similar (in terms of vertex degree) to their co-authors.

%% file: threats.tex
\section{Threats to Validity}\label{sec:threats}
\noindent\textbf{Construct Validity.} Our
results depend on the accuracy of DOA calculations. 
Currently, we compute DOA values using weights 
from the analysis of other systems~\cite{Fritz2010, Fritz2014}.
Although the authors of the absolute DOA claim their weights are general, we 
cannot fully eliminate the threat that the choice of weights pose to our 
results. Still, we have previously applied them when analyzing different 
open-source systems, obtaining positive feedback from 
developers~\cite{Avelino2016}. 
\\[-.2cm]

\noindent\textbf{Internal Validity.} We measure authorship considering only the
commit history of the official Linux kernel Git repository. Hence, we 
do not consider forks that are not merged into the mainstream development. 
Although these changes might be relevant to some  (e.g., studies about 
integration activities~\cite{German2015}), 
they are not relevant when measuring authorship of the official Linux kernel 
codebase. We also consider that all commits have the same importance. 
As such, we do not account for the granularity of changes 
(number of lines of code affected) nor their 
semantics (e.g., bug fixes, new features, refactoring, etc).
\\[-.2cm]

\noindent\textbf{External Validity.} The metrics we use  can
be applied to any software repository under a version control system. Still, our
findings are very specific to the Linux kernel development.
Thus, we cannot assume that the findings about workload,
specialization, and collaboration are general. Nonetheless, 
we pave the road for further studies to validate our findings in the context of 
other systems.

%% file: related.tex
\section{Related Work}\label{sec:related_work}

\vspace{0.2cm}
\noindent\textbf{Code Authorship.} McDonald and Ackerman propose the
\aspas{Line 10 Rule}, one of the first and most used heuristics for expertise
recommendation~\cite{McDonald2000}. 
The heuristic considers that the last person who changes a file
is most likely to be ``the'' expert. 
Expertise Browser~\cite{Mockus2002b} and Emergent Expertise
Locator~\cite{Minto2007} are alternative implementations to the \aspas{Line 10
Rule}. The former uses the concept of experience atoms (EA) to give the value 
for each developer's activity and takes into account the amount of {\sc EAs} to 
quantify expertise. The latter refines the Expertise Browser approach by 
considering the relationship between files that change together.
Fine-grained algorithms that assign expertise based on the percentage of
lines a developer has last touched are used by Girba et al.~\cite{Girba2005} and
by Rahman and Devanbu~\cite{Rahman2011}.

\vspace{0.2cm}
\noindent\textbf{Social Network Analysis (SNA).} 
Research in this area use information from source code
repositories to build a social network, adopting different strategies to
create the links between developers. 
Fern{\'{a}}ndez et al.~\cite{Lopez-Fernandez2006} apply SNA, linking developers that perform 
commits to the same module, to study their relationship and collaboration patterns. 
Others rely on fine-grained relations, building networks connecting developers 
that change the same file~\cite{Yang2014, Meneely2008,Jermakovics2011,Bird2011}. 
Joblin et al.~\cite{Joblin2015} propose an even more fine-grained approach, connecting 
developers that change the same function in a source code. They claim 
that file-based links result in dense networks, which obscures important 
network properties. 
Our approach, although centered on file-level information, does not produce dense 
networks, as authorship requires that developers make 
significant contributions to a file. 

%% file: conclusion.tex
\section{Conclusion}\label{sec:conclusion}

In this paper, we extract and analyze authorship 
parameters from a successful case: the Linux kernel. By mining over 11 years of 
the Linux kernel commit history, we investigate how authorship changes over 
time, deriving
measures that other communities mirroring the Linux kernel evolution could
directly replicate. Moreover, our study provides the grounds for further 
analyses---we define authorship concepts setting basic terminology and 
operationalization, in addition to providing a dataset of a large case study that others may use as a comparison baseline.
As future work, we seek to validate our findings directly with Linux kernel
developers. Moreover, we plan to study authorship in other systems.

\section*{Acknowledgment}
This study is supported by grants from FAPEMIG, CA\-PES, CNPq, and UFPI.

%% file: oss2017.bbl
\begin{thebibliography}{10}
\providecommand{\url}[1]{\texttt{#1}}
\providecommand{\urlprefix}{URL }

\bibitem{Barabasi2002a}
Albert, R., Barab\'asi, A.L.: Statistical mechanics of complex networks.
  Reviews of Modern Physic  74,  47--97 (2002)

\bibitem{Avelino2016}
Avelino, G., Passos, L., Hora, A.C., Valente, M.T.: A novel approach for
  estimating truck factors. In: 24th International Conference on Program
  Comprehension (ICPC). pp. 1--10 (2016)

\bibitem{designrules}
Baldwin, C.Y., Clark, K.B.: Design rules: the power of modularity. MIT Press
  (1999)

\bibitem{Bettenburg2015}
Bettenburg, N., Hassan, A.E., Adams, B., German, D.M.: Management of community
  contributions. Empirical Software Engineering  20(1),  252--289 (2015)

\bibitem{Bird2011}
Bird, C., Nagappan, N., Murphy, B., Gall, H., Devanbu, P.: Don't touch my code!
  In: 19th International Symposium on Foundations of Software Engineering
  (FSE). pp. 4--14 (2011)

\bibitem{chacon2014pro}
Chacon, S., Straub, B.: Pro Git. Expert's voice in software development,
  Apress, 2nd edn. (2014)

\bibitem{Corbet2013}
Corbet, J., Kroah-Hartman, G., McPherson, A.: {Who writes Linux: Linux kernel
  development: how fast it is going, who is doing it, what they are doing, and
  who is sponsoring it}. Tech. rep., Linux Foundation (2013),
  \url{http://www.linuxfoundation.org/publications/linux-foundation/who-writes-linux-2013}

\bibitem{Corbet2005}
Corbet, J., Rubini, A., Kroah-Hartman, G.: Linux device drivers. O'Reilly, 3rd
  edn. (2005)

\bibitem{Fritz2014}
Fritz, T., Murphy, G.C., Murphy-Hill, E., Ou, J., Hill, E.:
  {Degree-of-knowledge: modeling a developer's knowledge of code}. ACM
  Transactions on Software Engineering and Methodology  23(2),  14:1--14:42
  (2014)

\bibitem{Fritz2010}
Fritz, T., Ou, J., Murphy, G.C., Murphy-Hill, E.: A degree-of-knowledge model
  to capture source code familiarity. In: 32nd International Conference on
  Software Engineering (ICSE). pp. 385--394 (2010)

\bibitem{German2015}
German, D.M., Adams, B., Hassan, A.E.: {Continuously mining distributed version
  control systems: an empirical study of how Linux uses Git}. Empirical
  Software Engineering  (2015)

\bibitem{Gini}
Gini, C.: Measurement of inequality of incomes. The Economic Journal  31(121),
  124--126 (1921)

\bibitem{Girba2005}
Girba, T., Kuhn, A., Seeberger, M., Ducasse, S.: How developers drive software
  evolution. In: 8th International Workshop on Principles of Software Evolution
  (IWPSE). pp. 113--122 (2005)

\bibitem{Herbsleb2007}
Herbsleb, J.D.: {Global software engineering: the future of socio-technical
  coordination}. In: 2007 Future of Software Engineering (FOSE). pp. 188--198
  (2007)

\bibitem{Hubert2008}
Hubert, M., Vandervieren, E.: An adjusted boxplot for skewed distributions.
  Computational Statistics and Data Analysis  52(12),  5186--5201 (2008)

\bibitem{Jermakovics2011}
Jermakovics, A., Sillitti, A., Succi, G.: {Mining and visualizing developer
  networks from version control systems}. In: 4th International Workshop on
  Cooperative and Human Aspects of Software Engineering (CHASE). pp. 24--31
  (2011)

\bibitem{Joblin2015}
Joblin, M., Mauerer, W., Apel, S., Siegmund, J., Riehle, D.: From developer
  networks to verified communities: a fine-grained approach. In: 37th
  International Conference on Software Engineering (ICSE). pp. 563--573 (2015)

\bibitem{Lopez-Fernandez2006}
L{\'{o}}pez-Fern{\'{a}}ndez, L., Robles, G., Gonzalez-Barahona, J.M., Herraiz,
  I.: Applying social network analysis techniques to community-driven libre
  software projects. International Journal of Information Technology and Web
  Engineering pp. 27--48 (2006)

\bibitem{McDonald2000}
McDonald, D.W., Ackerman, M.S.: {Expertise recommender: a flexible
  recommendation system and architecture}. In: Conference on Computer Supported
  Cooperative Work (CSCW). pp. 231--240 (2000)

\bibitem{Meneely2011}
Meneely, A., Williams, L.: Socio-technical developer networks: Should we trust
  our measurements? In: 33rd International Conference on Software Engineering
  (ICSE). pp. 281--290 (2011)

\bibitem{Meneely2008}
Meneely, A., Williams, L., Snipes, W., Osborne, J.: Predicting failures with
  developer networks and social network analysis. In: 16th International
  Symposium on Foundations of Software Engineering (FSE). pp. 13--23 (2008)

\bibitem{Minto2007}
Minto, S., Murphy, G.C.: Recommending emergent teams. In: 4th Workshop on
  Mining Software Repositories (MSR). pp. 5--5 (2007)

\bibitem{Mistrik2010}
Mistr\'{i}k, I., Grundy, J., van~der Hoek, A., Whitehead, J.: Collaborative
  software engineering: challenges and prospects, pp. 389--403. Springer (2010)

\bibitem{Mockus2002b}
Mockus, A., Herbsleb, J.D.: {Expertise browser: a quantitative approach to
  identifying expertise}. In: 24th International Conference on Software
  Engineering (ICSE). pp. 503--512 (2002)

\bibitem{Newman2004}
Newman, M.E.J.: Coauthorship networks and patterns of scientific collaboration.
  Proceedings of the National Academy of Sciences  101,  5200--5205 (2004)

\bibitem{Newman2003}
Newman2, M.E.J.: Mixing patterns in networks. Physical Review E  67,  026126
  (2003)

\bibitem{Parnas1972}
Parnas, D.L.: On the criteria to be used in decomposing systems into modules.
  Communications of the ACM  15(12),  1053--1058 (1972)

\bibitem{Passos2015}
Passos, L., Padilla, J., Berger, T., Apel, S., Czarnecki, K., Valente, M.T.:
  {Feature scattering in the large: a longitudinal study of Linux kernel device
  drivers}. In: 14th International Conference on Modularity. pp. 81--92 (2015)

\bibitem{Rahman2011}
Rahman, F., Devanbu, P.: Ownership, experience and defects. In: 33rd
  International Conference on Software Engineering (ICSE). pp. 491--500 (2011)

\bibitem{sullivan-dsm}
Sullivan, K.J., Griswold, W.G., Cai, Y., Hallen, B.: The structure and value of
  modularity in software design. In: 9th International Symposium on Foundations
  of Software Engineering (FSE). pp. 99--108 (2001)

\bibitem{Watts1998}
Watts, D.J., Strogatz, S.H.: Collective dynamics of small-world networks.
  Nature  393,  440--2 (1998)

\bibitem{Yang2014}
Yang, X.: Social network analysis in open source software peer review. In: 22nd
  International Symposium on Foundations of Software Engineering (FSE). pp.
  820--822 (2014)

\end{thebibliography}
